\documentclass[aps,prb,twocolumn]{revtex4}
\usepackage{tabularx,graphicx}
\usepackage{amsmath, amsthm, amssymb} 
\usepackage{bbm}
\newcommand{\K}{\bf K}
\renewcommand{\r}{{\bf r}}
\begin{document}
\title{Few-electron physics in a nanotube quantum dot with spin-orbit coupling}
\author{B. Wunsch}
\affiliation{Department of Physics, Harvard University, Cambridge, Massachusetts 02138, USA}
\date{\today}
\email{bwunsch@physics.harvard.edu}
\begin{abstract}
We study the few-electron eigenspectrum of a nanotube quantum dot with spin-orbit coupling.  
The two-electron phase diagram as a function of the length of the dot and the applied parallel magnetic field shows clear signatures of both spin-orbit coupling and electron-electron interaction. 
Below a certain critical length, ground state transitions are correctly predicted by a single-particle picture and are mainly independent of the length of the dot despite the presence of strong correlations.
However, for longer quantum dots the critical magnetic field strongly decreases with increasing length, which is a pure interaction effect. In fact, the new ground state is spin- and valley-polarized, which implies a strong occupation of higher longitudinal modes. 
\end{abstract}
\maketitle

\section{Introduction}
Carbon nanotubes have allowed to realize clean quasi-one-dimensional electron systems.
Experiments revealing fundamental interaction effects include the detection of Wigner crystallization\cite{Deshpande08} or Luttinger liquid like behavior\cite{Bockrath99,Egger97}. 
An interesting feature of nanotubes is that the orbital part of low lying excitations has an additional spin-like degree of freedom the valley index.
This new degree of freedom can cause orbital Kondo effect\cite{Pablo05}  unusual spin configurations\cite{Oreg00}, or a new type of shell structure \cite{Moriyama05}  in nanotube quantum dots. 

It was generally assumed that spin and valley degrees of freedom lead to a fourfold degeneracy of electronic states, however, recently spin-orbit coupling was observed to split this degeneracy in two pairs of either parallel and antiparallel spin and valley orientation\cite{Kuemmeth08}.  
Interestingly the experimental data could be well explained in a single-particle picture, and correlation effects seemed to be of minor importance.
In this work we analyze how interaction effects show up in the two-particle spectrum of a single nanotube quantum dot with spin-orbit coupling. We argue that  the eigenspectrum can be divided in multiplets of states that have the same orbital symmetry.  Energy gaps within the same multiplet are only determined by spin-orbit coupling and the orbital Zeeman effect (and additional small correction due to local interactions), and are therefore captured in a single-particle picture. However, the extend of correlations can be appreciated by comparing different multiplets. 
In particular we show that above a certain critical length a tiny magnetic field is enough to cause a ground state transition to a spin and valley-polarized two-particle state, that necessarily involves the occupation of higher modes.

In the next section we introduce our model. The quantum dot is described by a potential well along the nanotube and a continuum description is applied for the single-particle spectrum of electrons localized in this well and subject to a parallel magnetic field\cite{Bulaev08}. The single-particle spectrum also includes the effect of spin-orbit coupling\cite{Kuemmeth08, Paco06}. We then show how the electron-electron interaction can be correctly incorporated in the continuum model\cite{Egger97,Odintsov99}.
Thereafter we present our results, including a detailed discussion of the phase diagram of the two-electron ground state as a function of magnetic field and length of the quantum dot.

\section{Model}
A nanotube is a one-atom-thick layer of graphite called graphene wrapped into a seamless cylinder. Depending on the orientation of the underlying honeycomb lattice of
carbon atoms with respect to the symmetry axis of the nanotube, it is either metallic or semiconducting\cite{Ando05}. We will study a semiconducting nanotube
with an additional confinement potential along the tube, which is controlled by external gates and gives rise to a discrete set of localized electronic states.
 
\subsection{Single-particle spectrum}

In a continuum description, the single
particle orbitals have two components belonging to the two
sublattices, called $A$ and $B$ in the following. Furthermore, the single-particle states have an additional
spin-like degree of freedom $\tau\in \pm 1$, the valley index, since
there are two inequivalent band minima at the ${\bf K}$ and ${\bf K'}=-{\bf K}$ points of
the graphene's Brillouin zone.

Using cylindrical coordinates $\zeta,\phi$ the single-particle Hamiltonian is given by
\begin{eqnarray}
  H_0&=&-i \hbar v_F (\tau \sigma_x \frac{1}{R}\partial_\phi + \sigma_y \partial_\zeta) +V(\zeta)\,,
\label{eq:H0}
\end{eqnarray}
where $v_F$ is the Fermi velocity and $\sigma_x,\sigma_y$ are Pauli
matrices acting on the sublattice space.  We study a square well potential, i.e. $V(\zeta)$ is zero for $|\zeta|<L/2$ and $V_G$
otherwise\cite{Bulaev08}. We assume the potential to be smooth on
the atomic length scale (interatomic distance $a_0=a/\sqrt{3}=0.142$ nm, where $a$ is the lattice spacing) and
therefore neglect confinement induced intervalley scattering.  

The single
particle solutions are given by:
\begin{eqnarray}
\Psi_{\tau \kappa k}(\r)=(2\pi R)^{-1/2} e^{i\tau \K \r} e^{i \kappa R \phi} \phi_k(\zeta)\,,\label{eq:WF}
\end{eqnarray}
where $k, \kappa$ denote the wavevectors along and around the tube and the two component longitudinal wavefunction is normalized such that $\int d\zeta\,( |\phi_{A k}|^2(\zeta)+|\phi_{B k}|^2(\zeta))=1$. 
$ \phi_k(\zeta)$ is given by a
standing wave with wavevector $k$ inside the well and  evanescent modes outside the well\cite{Bulaev08}. The corresponding eigenenergy is given by $E_k=\hbar v_F \sqrt{\kappa^2+k^2}$. We note that we measure energy with respect to the center of the gap, so that the dominant part of the single-particle energy is constant and given by $\hbar v_F \kappa \approx 220 meV/R[nm]$. Electron-electron interaction however, affect the longitudinal part with a much smaller level spacing that depending on the length of the dot is $\Delta=2-10$ meV.

Both the axial magnetic field $B$  and the spin-orbit coupling
modify the transverse wavevector $\kappa$
\begin{equation}
\kappa=\tau/R(1/3+\tau \Phi/\Phi_0+\tau \sigma \Phi_{SO}/\Phi_0)\,.\label{eq:kx}
\end{equation}
Here $\sigma$ denotes the spin component along the tube, $\Phi=\pi R^2 B$ the magnetic flux through the tube, $\Phi_0=h/e$, and $\Phi_{SO}\approx
7.2\;\, 10^{-4}$ determines the curvature induced spin-orbit
interaction\cite{Paco06,Kuemmeth08}.  The second term in Eq.~(\ref{eq:kx}) results from the coupling between the external magnetic field and the orbital magnetic moment that is caused by the transverse motion around the tube\cite{Minot04}. Electrons in different valleys have an opposite sign of this orbital momentum, which leads to a valley splitting that is linear in the applied magnetic field. We call this the orbital Zeeman term in analogy with the smaller spin Zeeman term $H_Z=-g\mu_B \sigma B/2$ that leads to spin dependent energy shift in the magnetic field.
The orbital magnetic moment is $\mu_{orb}\approx -\tau\,0.5\, R[\text{nm}]$ meV/T and the spin magnetic moment $\mu_{spin} \approx \sigma\,0.06$ meV/T . The third term  in Eq.~(\ref{eq:kx}) describes spin-orbit coupling.  It increases (decreases) the energy of single-particle states with aligned (anti-aligned) spin and valley degree of freedom. The energy splitting is approximately given by $\Delta_{SO}\approx 1.9/d[\text{nm}]$ meV, where $d=2R$ denotes the diameter of the tube.\cite{Kuemmeth08}  

Spin-orbit coupling and orbital Zeeman effect couple to the transverse part of the wavefunction while their effect on the longitudinal part $\phi_k(\zeta)$ can be neglected for the large dot sizes we are interested in.
The longitudinal wavevector $k$ is determined by the transcendental
equation\cite{Bulaev08}
\begin{eqnarray}
\tan(k L)=\frac{\tilde{k} k}{E_k (E_k-V_G)/(\hbar v_F)^2 -\kappa^2}\,,
\end{eqnarray}
where $\tilde{k}=\left(\kappa^2-\left[(E_k-V_G)/\hbar
    v_F\right]^2\right)^{1/2}$ determines the decay of the wavefunction outside
the well.  Due to the symmetries of the Hamiltonian in Eq.~(\ref{eq:H0}), $\phi_k(\zeta)$ is
real and has a well-defined parity $p=\pm1$, $\phi_A(\zeta)=p \phi_B(-\zeta)$, where
$A$, $B$ label the two sublattices. The parity of the $i$-th mode (where the ground state corresponds to $i=0$) is given by $p=(-1)^i$.

\begin{figure}
  \begin{center}
    \includegraphics*[angle=0,width=0.7\linewidth]{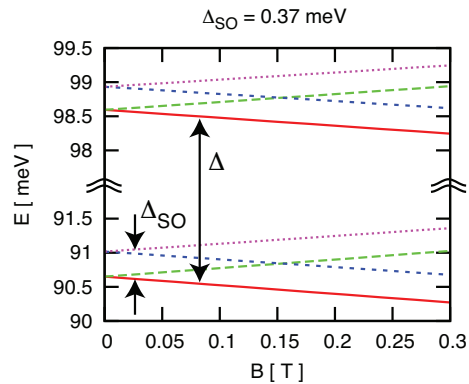} 
    \caption{(Color online) Single-particle states for a $R=2.5$nm nanotube with a
      $L=80$nm square well of depth $V_g=50meV$.  Color coding:
      Red (solid): $\tau=-$, $\sigma=\uparrow$, Green (long dashed):  $\tau=+1$ $\sigma=\downarrow$, Blue (short dashed): $\tau=-$, $\sigma=\downarrow$, Purple (dotted): $\tau=+$, $\sigma=\uparrow$.}
\label{fig:N1_SP}
\end{center}
\end{figure}

Figure~\ref{fig:N1_SP} shows the magnetic field dependence of the two lowest longitudinal modes. Each
mode gives rise to four single-particle states due to the two spin and
two valley degrees of freedom. At zero magnetic field these four
states are split in two Kramer doublets (states obtained by flipping
simultaneously spin and valley degree of freedom are degenerate
due to time-reversal symmetry).  Finite magnetic fields lead to energy
shifts linear in magnetic field caused by orbital and spin Zeeman
splitting. Denoting the single-particle states by $P,\tau,\sigma$, a single-particle picture predicts the two-particle ground state  to be $| {+1} , -, \uparrow;\, {+1}, +, \downarrow\,\rangle$ for $B<B_{crit}\approx0.15T$ and
$| {+1} , -, \uparrow;\, {+1}, -, \downarrow\,\rangle$ for $B>B_{crit}$. We note that only for ridiculously large magnetic fields of about 80 Tesla the single-particle picture predicts a further ground state crossing due to an occupation of the first excited shell  $| {+1} , -, \uparrow;\, {-1}, -, \uparrow\,\rangle$.

\subsection{Interaction}

Due to the large energy gap between different transverse modes, it is justified to treat completely filled as well as completely empty transverse subbands as inert, giving rise to a static screening constant $\epsilon$. 
Assuming gate electrodes to be sufficiently far
away from the quantum dot, we use a long-ranged interaction between the conduction
electrons $U_I(\r_1,\r_2)=e^2/(\epsilon |\r_1-\r_2|)$.

The interaction does not depend on the electron spin and is
therefore diagonal in the spin degree of freedom. 
However, the local part of the interaction is not diagonal in the valley degree of freedom\cite{Egger97, Odintsov99}. 
It is therefore instructive to split the interaction $H_{int}$ in a long-ranged part $V_C$ and a local, onsite interaction $V_H$.

After integrating out the transverse motion the interaction is given by: 
\begin{eqnarray*}
H_{int}&=&V_C+V_H\\
V_C&=&\frac{1}{2}\int\,d\zeta_1\,d\zeta_2 V(\zeta_1-\zeta_2) \rho(\zeta_1)\rho(\zeta_2)\\
V_H&=&V_H^{(1)}+V_H^{(2)}\\
V_H^{(1)}&=&\tilde{U} \sum_{p,\tau_1,\tau_2} \int\,d\zeta\, \psi^\dag_{p \uparrow \tau_1}(\zeta) \psi^\dag_{p \downarrow \tau_2}(\zeta) \psi_{p \downarrow \tau_2}(\zeta) \psi_{p \uparrow \tau_1}(\zeta) \\
V_H^{(2)}&=&\tilde{U} \sum_{p,\tau} \int\,d\zeta \psi^\dag_{p \uparrow \tau}(\zeta) \psi^\dag_{p \downarrow \bar{\tau}}(\zeta) \psi_{p \downarrow \tau}(\zeta) \psi_{p \uparrow \bar{\tau}}(\zeta) \\
\end{eqnarray*}
Here $\rho(\zeta)=\sum_{p\,\sigma\,\tau} \psi^\dag_{p \sigma \tau}(\zeta) \psi_{p \sigma \tau}(\zeta)$ labels the charge density at $\zeta$, where $p\in\{A,B\}$ denotes the sublattice index.
The field operators are now expressed by $\psi_{p \sigma \tau}(\zeta)=\sum_{k}\phi_{p k}(\zeta) a_{\sigma \tau k}$ where $a_{\sigma \tau k}$  annihilates an electron  in the longitudinal mode $k$ (characterized by the wavefunction $\phi_k$ of Eq.~(\ref{eq:WF})) and with spin $\sigma$ and valley $\tau$.  
The now one-dimensional interaction $V(\zeta)=2 e^2 K[4R^2/(\zeta^2+4 R^2)]  /[\epsilon \pi(\zeta^2+4 R^2)^{1/2}]$ can be expressed by the incomplete elliptical integral of first kind K(x) \cite{abramowitz}. 
In the following the strength of the long-ranged interaction is characterized by the effective fine structure constant $\alpha=e^2/\epsilon \hbar v_F\approx 2.2/\epsilon$.
The local part depends on 
$\tilde{U}= U A_u/( 2\pi R)$, where $A_u=a^2 \sqrt{3}/2$ denotes the size of the graphene's unit cell in real space and $U$ is the onsite interaction. We use $U=15$eV.\cite{Oreg00}

Due to the rapidly oscillating Bloch factors $e^{i\tau \K \r}$ of the
eigenfunctions (\ref{eq:WF}), the long-ranged interaction $V_C$ is diagonal in the valley and spin degrees of freedom, and in agreement with the continuum description, the interatomic distance between the two sublattices is neglected. The lattice effects not captured in the continuum model and the long-ranged interaction $V_C$ are taken into account by the local part of the interaction $V_H$. 
We note that $V_C$ depends equally on spin- and valley symmetry, but this is not the case for local interaction.
For example spin-aligned electrons do not interact via $V_H$, but valley-aligned do. While local interactions can be very important for short quantum dots\cite{Kostyrko08} their effect is rather small for the long quantum dots as shown in the following. However, also the spin-orbit coupling is a small quantity and as we will discuss below local interaction energies can add up to the  spin-orbit interaction.
We note that $V_H$ still  conserves valley polarization $\sum_i \tau_i$ and that it allows for intervalley exchange interaction ($V_H^{(2)}$).

The many body eigenfunctions can be characterized by a triple of quantum numbers $(P,S_z,T_z)$, where  
$P=\prod_n p_n \in\{\pm1\}$ denotes the total parity, $S_z=1/2 \sum_n \sigma_n$ the z- component of the total spin and $T_z=1/2 \sum_n \tau_n$ the total valley polarization; where $n=1,..,N_e$ runs over all electrons. 
We note that without local and spin-orbit interaction the two-particle states can also be chosen as eigenstates of total spin $S^2$ and total valley degree of freedom $T^2$.

In the following we calculate the few-electron eigenspectrum of the Hamiltonian $H=H_0+H_Z+H_{int}$.
We restrict the single-particle basis to the bound longitudinal modes of the lowest transverse mode and diagonalize the few-electron Hamiltonian for each set of conserved quantum numbers $(P, T_z, S_z)$, so that electron-electron correlations within this basis set are fully taken into account.

\section{Results}

\begin{figure}
  \begin{center}
    \includegraphics*[angle=0,width=\linewidth]{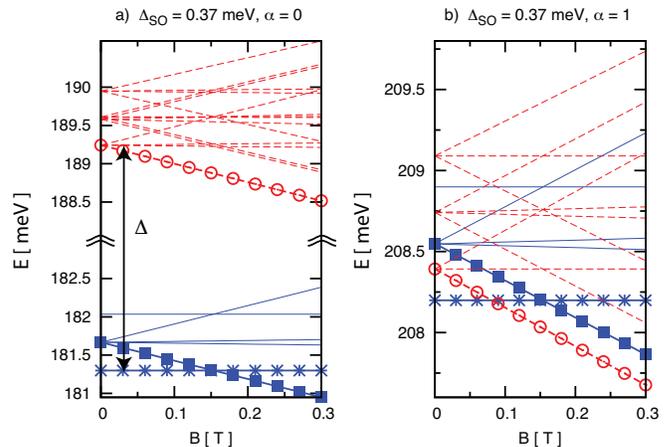}
    \caption{(Color online) Energies of lowest two-particle states. Left no interaction, right
    interaction $\alpha=e^2/\epsilon \hbar v_F=1$. Parameters as in
    Fig~\ref{fig:N1_SP}.  Blue (solid): Parity $+1$, Red (dashed): Parity $-1$. Three
    states are marked:Crosses: $(P=1,T_z=0,S_z=0)$,  Filled squares: $(1, -1,0)$, 
     Open circles: $(1,-1, 1)$.}
\label{fig:N2_B}
\end{center}
\end{figure}

\begin{figure}
  \begin{center}
    \includegraphics*[angle=0,width=\linewidth]{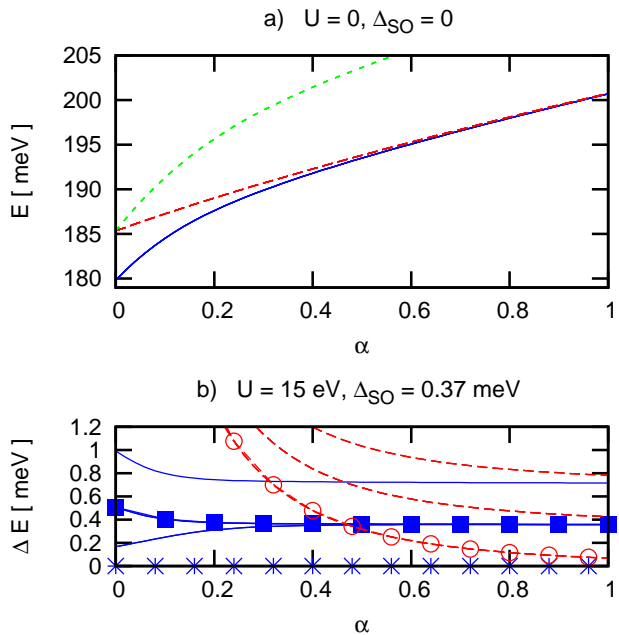}
    \caption{(Color online) Dependence of two-electron spectra on strength of long-ranged interaction $\alpha$ in absence of a magnetic field for $R=2.5$nm, $L=100$nm. a) Total two-particle energies in absence of local interactions and spin-orbit coupling. Colors label symmetry of the longitudinal part of the wavefunction. Blue (solid) and green (short dashed) indicate symmetric and red (long dashed) an antisymmetric longitudinal wavefunction.  b) Two-particle excitation energy $\Delta E=E_i(N=2)-E_0(N=2)$. Energy splitting within each multiplet is determined by spin-orbit coupling and local interactions.  Meaning of marked states as in FIG.~\ref{fig:N2_B}.}
\label{fig:N2_B0_Vc}
\end{center}
\end{figure}
Figure~\ref{fig:N2_B} shows the two-particle spectrum as a function of magnetic field both for non-interacting (left part $\alpha=0$) and interacting electrons (right part $\alpha=1$).
For non-interacting electrons the ground state corresponds to a double-occupation of the lowest longitudinal mode and has parity $P=+1$ (blue states).  
Since the spacing $\Delta$ to the next longitudinal mode is much larger than the spin-orbit splitting, states with negative parity $P=-1$ (red states) are energetically well separated from the ground state for all relevant magnetic fields. In the following we label eigenstates by their quantum numbers $(P,T_z,S_z)$.
Without magnetic field the nondegenerate ground state corresponds to the subspace with $(1, 0,  0)$ (blue line with crosses) which is favored by spin-orbit coupling.
At a critical magnetic field the ground state crosses to $(1, -1, 0)$ (blue line with filled squares) due to the orbital Zeeman term. 

Electron-electron interaction strongly reduces the gap between the $P=1$ and $P=-1$ states as shown on the right side of FIG.~\ref{fig:N2_B}. At the same time spin-orbit induced energy gaps are unaffected by interactions and also the magnetic field dependence of the energies is the same as in the noninteracting case.
For the parameters chosen in FIG.~\ref{fig:N2_B}  the ground state transition at finite magnetic field occurs to the  $(-1,-1,1)$  state, which is spin and valley-polarized (red line with open circles in FIG.~\ref{fig:N2_B}).

We now study the reduction of the energy spacing between the $P=+1$ and $P=-1$ multiplet of states with increasing interactions. 
Since the magnetic field dependence of the energies is hardly changed by interactions it is instructive to study the spectrum in absence of magnetic fields. 
We first neglect spin-orbit coupling, $\Delta_{SO}=0$ and the onsite interaction, $U=0$.
Figure~\ref{fig:N2_B0_Vc}~a) shows the eigenspectrum of $H_0+V_C$ at $B=0$ as function of the interaction strength $\alpha$. The blue (solid) line is the 6-fold degenerate $P=+1$ ground state. Interaction split the  $P=-1$ states in two sets. The red (long dashed) line consists  of $10$ states that approach the $P=+1$ ground state with increasing interactions. The green (short dashed) line indicates the remaining six states with $P=-1$. 
$V_C$ exclusively acts on the longitudinal part of the wavefunction since a single transverse subband is considered. The two-particle eigenstates can be factorized in longitudinal, spin and valley parts. The longitudinal part of  the $P=+1$ ground state is symmetric with respect to interchange of two electrons. 
It is multiplied with either valley-triplet and spin-singlet or valley-singlet and spin-triplet, in order to guarantee the antisymmetry of the total two-particle wavefunction. This explains the six-fold degeneracy.
The energetically favoured set of $P=-1$ states is ten-fold degenerate and has an antisymmetric orbital part and spin and valley part are either both singlet or both triplet.
Figure~\ref{fig:N2_B0_Vc} a) shows that the two-particle ground state has always a symmetric orbital part for all interaction strengths, however the energy gap to the eigenstates with antisymmetric longitudinal part vanishes with increasing interaction. We note that for scalar eigenfunctions of a Schr\"odinger equation a symmetric orbital part is guaranteed by the Lieb-Mattis theorem\cite{Lieb61}.
Since we are describing a semiconducting nanotube with a large gap between transverse modes, we are in fact very close to that limit. With increasing $\alpha$, electrons become more and more correlated and finally form a quasi classical Wigner crystal, where electrons are localized in different region of space and symmetry becomes irrelevant.

Including again spin-orbit coupling and local interactions states with an originally symmetric (antisymmetric) longitudinal part split in multiplets of six (ten) states.
This splitting is small with respect to the total energy, so that Figure~\ref{fig:N2_B0_Vc} b) shows the excitation energy $\Delta E=E_i(N=2)-E_0(N=2)$ rather than the total energies.
The two-electron ground state, which belongs to $(+1,0,0)$, therefore defines $\Delta E=0$ axis. Excitations to $P=+1$ ($P=-1$) states are depicted in blue (red).
Since an increase of the interaction strength $\alpha$ leads to an increasing distance between the two electrons the probability of finding both on the same site strongly decreases with increasing $\alpha$ and the local interaction effects vanish. The energy splitting within the different multiplets therefore quickly approaches the constant spin-orbit gap with increasing $\alpha$. 
Effects of local interactions are however visible if the quantum dot becomes shorter, or if $\alpha$ is small (we assume that local interactions are not screened). Onsite interactions favor spin triplet states over spin singlet states and increase the energy gap between the $(1,0,0)$ ground state, that is in a superposition of spin singlet and spin triplet, and
the $(1,-1,0)$ state (filled squares in FIG~\ref{fig:N2_B0_Vc} b)), which is a spin singlet state.

Above a critical magnetic field, a valley-polarized ground state ($T_z=-1$) is favored, the parity of which  depends on the ratio of single-particle and interaction energy. This ratio increases for decreasing length of the quantum dot or increasing radius of the nanotube or decreasing dielectric constant $\epsilon$. The length of the quantum dot is tunable experimentally by changing gate voltages.

\begin{figure}
  \begin{center}
    \includegraphics*[angle=-90,width=\linewidth]{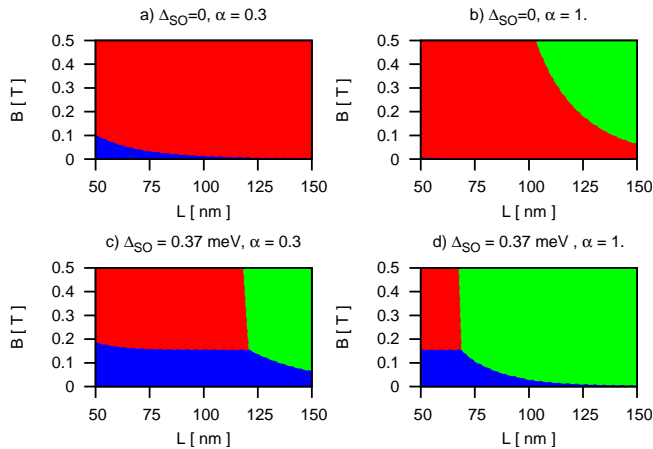}
    \caption{(Color online) Two-particle ground state as a function of magnetic field
      and length of the nanotube for $R=2.5$nm. The ground state belongs to one of the following sets of quantum numbers:
       Blue: $(1,0,0)$  Red: $(1,-1,0)$,
       Green: $(-1,-1,1)$.
      Upper (lower) row shows phase diagrams in absence (presence) of spin-orbit coupling and left
      (right) column shows different interaction strengths of long-ranged Coulomb interaction.}
\label{fig:N2_PD}
\end{center}
\end{figure}

\begin{figure}
  \begin{center}
    \includegraphics*[angle=-90,width=0.6\linewidth]{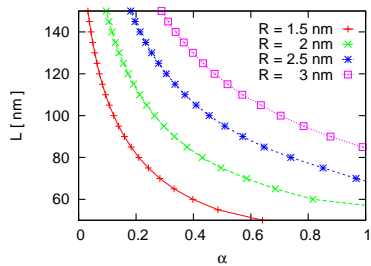}
    \caption{(Color online) Dependence of critical length $L$ on interaction strength $\alpha$  and radius $R$ of the tube. If the quantum dot is longer than the critical length the spin and valley-polarized state $(-1,-1,1)$ becomes ground state at a small magnetic field. Spin-orbit coupling $\Delta_{SO}=1.9 \text{ meV}/d$[nm].}
\label{fig:N2_VcCrit}
\end{center}
\end{figure}

\begin{figure}
  \begin{center}
    \includegraphics*[angle=-90,width=\linewidth]{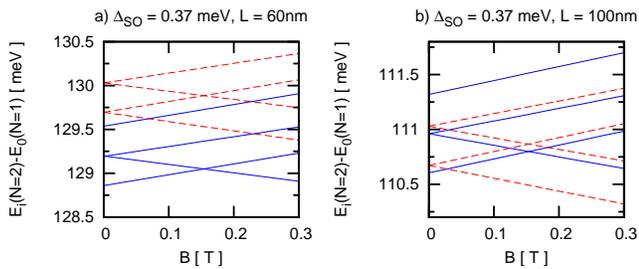}
    \caption{(Color online) Energy needed to add a second particle to the quantum dot. Only transitions that are allowed by spin and valley selection rules are shown.  $R=2.5$nm, $\alpha=1$. }
     \label{fig:N2_AddEn}
\end{center}
\end{figure}

We now discuss the phase diagram of the two-electron quantum dot as a function of magnetic field $B$, and length $L$ of the quantum dot, for different spin-orbit couplings $\Delta_{SO}$ and interaction strengths $\alpha$ as shown in FIG.~\ref{fig:N2_PD}. We note that the appearance of the spin and valley-polarized state (green areas) is favored by the interplay between spin-orbit coupling and long-ranged Coulomb interaction.
Figure~\ref{fig:N2_PD} a) and b) show the phase diagram without spin-orbit coupling.  Then the ground state in zero field is given by the three spin-polarized states (among which the  $(1,0,0)$ one is indicated by the blue area on Fig \ref{fig:N2_PD} ). They are separated from the three spin singlet states of the $P=1$ multiplet by local interaction effects only. At a critical magnetic field the valley-polarized state $(1,-1,0)$ (red area of Fig \ref{fig:N2_PD}) is favored due to the orbital Zeeman term. In agreement with our discussion above, the local interaction is more relevant for shorter quantum dots and for small $\alpha$. For sufficiently large quantum dots the $(-1,-1,1)$ state (green area of Fig \ref{fig:N2_PD}) becomes the ground state since long-ranged Coulomb interaction strongly suppresses the level spacing between the $P=1$ and $P=-1$ states and the remaining gap can be compensated by the gain in the (spin) Zeeman term.

Figure~\ref{fig:N2_PD} c) and d)  show that the phase diagram drastically changes if spin-orbit coupling is included. 
The zero-field ground state (still belonging to $(1,0,0)$) is now non-degenerate and is in a superposition of spin singlets and triplets. Additionally, the regions where the ground state belongs to either $(1,0,0)$ or $(-1,-1,1)$ are both considerably enlarged at the expense of the $(1,-1,0)$ state.
Below a critical length, the magnetic field where the ground state crossing (from $(1,0,0)$ to $(1,-1,0)$) occurs is mostly length independent and coincides with the value predicted by a single-particle picture\cite{Kuemmeth08}. In contrast for somewhat larger quantum dots the transition occurs to  the $(-1,-1,1)$ state
and the corresponding critical magnetic fields decrease continuously with increasing length of the quantum dot.

In the phase diagrams of the two-particle ground state with spin-orbit coupling there is generally a critical length above which a magnetic field causes a ground state transition to the spin and valley-polarized state. The dependence of this critical length on $\alpha$ and the radius of the nanotube is shown in FIG.~\ref{fig:N2_VcCrit}. In agreement with our discussion this critical length decreases with increasing interaction strength or decreasing radius.

A powerful tool to measure the few-electron spectrum of the quantum dot is transport spectroscopy for different lengths of the quantum dot.
Such an experiment allows to measure the energy needed to cause a transition from the one-electron ground state of energy $E_0(N=1)$ to a two-particle excited state $E_i(N=2)$ where $i$ denotes the excitation.
Allowed transitions from the one-particle ground state to a two-particle excited state cannot change $T_z$ or $S_z$ by more than $\pm 1/2$. These transitions are depicted in FIG.~\ref{fig:N2_AddEn}. For the shorter quantum dot, the energy needed for the $N=1$ to $N=2$ ground state transition exactly follows the first excited one-electron energy (except for a constant charging energy). This is not the case for the longer quantum dot, where the two-electron ground state transition occurs at smaller fields. We note that excitations between two-particle states of different parity are always modified by interactions and are not a mere combination of level spacing and spin-orbit gaps.

\section{Conclusions}
We have presented a detailed study of the two-electron eigenspectrum of a nanotube quantum dot with spin-orbit coupling.
Generally we find that the eigentates are strongly correlated and by varying the length of the quantum dot we identify clear signatures of short and long-ranged interaction. 
In particular we studied the two-electron phase diagram as a function of the length of the quantum dot and the applied magnetic field. While the ground state at zero magnetic field always corresponds to the same set of  quantum numbers (given by parity $P=1$, spin $S_z=0$ and  valley polarization $T_z=0$) for all lengths,  a finite magnetic field causes a transition to a valley-polarized state which is either  a spin singlet or a spin triplet, depending on the length of the quantum dot. 
The former case is the one predicted by a single-particle picture since valley-polarized electrons in the lowest mode must be in a spin singlet state. This crossing is unaltered by Coulomb interaction, since the two crossing states have the same orbital part and their splitting is only given by the length independent orbital Zeeman shift and the spin-orbit gap (plus small correction due to local interactions).
However, once the length of the quantum dot exceeds a certain critical value interaction effects cause a ground state transition to a spin and valley-polarized state. Increasing the length even further, the magnetic field of the ground state transition becomes arbitrarily small.

An interesting continuation of our work is to analyze interaction effects on recently suggested optical manipulation schemes for the spin in carbon nanotubes\cite{Galland08}.  
Another open question is whether the presented interaction effects are also manifest in double quantum dots in nanotubes. In particular we suggest to study whether a valley and spin polarized ground state modifies selection rules leading to spin and/or valley blockade in transport spectroscopy\cite{Churchill08}.

{\it Note added}
While preparing this manuscript we became aware of the work of A. Secchi and M. Rontani~[\onlinecite{Secchi09}], who obtained similar results for a nanotube quantum dot with harmonic confinement instead of the potential well used here. We note that the agreement of general conclusions in both works shows the robustness of the discussed interaction effects. 

\section{Acknowledgements}
I am particularly thankful to Eugene Demler for heading me to this problem and for his guidance. Special thanks also to Ferdinand Kuemmeth, who gave me his experimental point of view and to Ana Maria Rey, Javier Stecher, Lars Fritz, Nicolaj Zinner and David Pekker for illuminating discussions. 
B. Wunsch is funded by the German Research Foundation.

\bibliography{SpinOrbit}

\begin{thebibliography}{18}
\expandafter\ifx\csname natexlab\endcsname\relax\def\natexlab#1{#1}\fi
\expandafter\ifx\csname bibnamefont\endcsname\relax
  \def\bibnamefont#1{#1}\fi
\expandafter\ifx\csname bibfnamefont\endcsname\relax
  \def\bibfnamefont#1{#1}\fi
\expandafter\ifx\csname citenamefont\endcsname\relax
  \def\citenamefont#1{#1}\fi
\expandafter\ifx\csname url\endcsname\relax
  \def\url#1{\texttt{#1}}\fi
\expandafter\ifx\csname urlprefix\endcsname\relax\def\urlprefix{URL }\fi
\providecommand{\bibinfo}[2]{#2}
\providecommand{\eprint}[2][]{\url{#2}}

\bibitem[{\citenamefont{Deshpande and Bockrath}(2008)}]{Deshpande08}
\bibinfo{author}{\bibfnamefont{V.~V.} \bibnamefont{Deshpande}}
  \bibnamefont{and} \bibinfo{author}{\bibfnamefont{M.}~\bibnamefont{Bockrath}},
  \bibinfo{journal}{nature physics} \textbf{\bibinfo{volume}{4}},
  \bibinfo{pages}{314} (\bibinfo{year}{2008}).

\bibitem[{\citenamefont{Bockrath et~al.}(1999)\citenamefont{Bockrath, Cobden,
  Lu, Rinzler, Smalley, Balents, and McEuen}}]{Bockrath99}
\bibinfo{author}{\bibfnamefont{M.}~\bibnamefont{Bockrath}},
  \bibinfo{author}{\bibfnamefont{D.~H.} \bibnamefont{Cobden}},
  \bibinfo{author}{\bibfnamefont{J.}~\bibnamefont{Lu}},
  \bibinfo{author}{\bibfnamefont{A.~G.} \bibnamefont{Rinzler}},
  \bibinfo{author}{\bibfnamefont{R.~E.} \bibnamefont{Smalley}},
  \bibinfo{author}{\bibfnamefont{L.}~\bibnamefont{Balents}}, \bibnamefont{and}
  \bibinfo{author}{\bibfnamefont{P.~L.} \bibnamefont{McEuen}},
  \bibinfo{journal}{Nature} \textbf{\bibinfo{volume}{397}},
  \bibinfo{pages}{598} (\bibinfo{year}{1999}).

\bibitem[{\citenamefont{Egger and Gogolin}(1997)}]{Egger97}
\bibinfo{author}{\bibfnamefont{R.}~\bibnamefont{Egger}} \bibnamefont{and}
  \bibinfo{author}{\bibfnamefont{A.~O.} \bibnamefont{Gogolin}},
  \bibinfo{journal}{Phys. Rev. Lett.} \textbf{\bibinfo{volume}{79}},
  \bibinfo{pages}{5082} (\bibinfo{year}{1997}).

\bibitem[{\citenamefont{Jarillo-Herrero
  et~al.}(2004)\citenamefont{Jarillo-Herrero, Kong, van~der Zant, Dekker,
  Kouwenhoven, and Franceschi}}]{Pablo05}
\bibinfo{author}{\bibfnamefont{P.}~\bibnamefont{Jarillo-Herrero}},
  \bibinfo{author}{\bibfnamefont{J.}~\bibnamefont{Kong}},
  \bibinfo{author}{\bibfnamefont{H.~S.} \bibnamefont{van~der Zant}},
  \bibinfo{author}{\bibfnamefont{C.}~\bibnamefont{Dekker}},
  \bibinfo{author}{\bibfnamefont{L.~P.} \bibnamefont{Kouwenhoven}},
  \bibnamefont{and} \bibinfo{author}{\bibfnamefont{S.~D.}
  \bibnamefont{Franceschi}}, \bibinfo{journal}{Nature}
  \textbf{\bibinfo{volume}{434}}, \bibinfo{pages}{484} (\bibinfo{year}{2004}).

\bibitem[{\citenamefont{Oreg et~al.}(2000)\citenamefont{Oreg, Byczuk, and
  Halperin}}]{Oreg00}
\bibinfo{author}{\bibfnamefont{Y.}~\bibnamefont{Oreg}},
  \bibinfo{author}{\bibfnamefont{K.}~\bibnamefont{Byczuk}}, \bibnamefont{and}
  \bibinfo{author}{\bibfnamefont{B.~I.} \bibnamefont{Halperin}},
  \bibinfo{journal}{Phys. Rev. Lett.} \textbf{\bibinfo{volume}{85}},
  \bibinfo{pages}{365} (\bibinfo{year}{2000}).

\bibitem[{\citenamefont{Moriyama et~al.}(2005)\citenamefont{Moriyama, Fuse,
  Suzuki, Aoyagi, and Ishibashi}}]{Moriyama05}
\bibinfo{author}{\bibfnamefont{S.}~\bibnamefont{Moriyama}},
  \bibinfo{author}{\bibfnamefont{T.}~\bibnamefont{Fuse}},
  \bibinfo{author}{\bibfnamefont{M.}~\bibnamefont{Suzuki}},
  \bibinfo{author}{\bibfnamefont{Y.}~\bibnamefont{Aoyagi}}, \bibnamefont{and}
  \bibinfo{author}{\bibfnamefont{K.}~\bibnamefont{Ishibashi}},
  \bibinfo{journal}{Phys. Rev. Lett.} \textbf{\bibinfo{volume}{94}},
  \bibinfo{pages}{186806} (\bibinfo{year}{2005}).

\bibitem[{\citenamefont{Kuemmeth et~al.}(2008)\citenamefont{Kuemmeth, Ilani,
  Ralph, and McEuen}}]{Kuemmeth08}
\bibinfo{author}{\bibfnamefont{F.}~\bibnamefont{Kuemmeth}},
  \bibinfo{author}{\bibfnamefont{S.}~\bibnamefont{Ilani}},
  \bibinfo{author}{\bibfnamefont{D.}~\bibnamefont{Ralph}}, \bibnamefont{and}
  \bibinfo{author}{\bibfnamefont{P.}~\bibnamefont{McEuen}},
  \bibinfo{journal}{Nature} \textbf{\bibinfo{volume}{452}},
  \bibinfo{pages}{448} (\bibinfo{year}{2008}).

\bibitem[{\citenamefont{Bulaev et~al.}(2008)\citenamefont{Bulaev, Trauzettel,
  and Loss}}]{Bulaev08}
\bibinfo{author}{\bibfnamefont{D.}~\bibnamefont{Bulaev}},
  \bibinfo{author}{\bibfnamefont{B.}~\bibnamefont{Trauzettel}},
  \bibnamefont{and} \bibinfo{author}{\bibfnamefont{D.}~\bibnamefont{Loss}},
  \bibinfo{journal}{Phys. Rev. B.} \textbf{\bibinfo{volume}{77}},
  \bibinfo{pages}{235301} (\bibinfo{year}{2008}).

\bibitem[{\citenamefont{Huertas-Hernando
  et~al.}(2006)\citenamefont{Huertas-Hernando, Guinea, and Brataas}}]{Paco06}
\bibinfo{author}{\bibfnamefont{D.}~\bibnamefont{Huertas-Hernando}},
  \bibinfo{author}{\bibfnamefont{F.}~\bibnamefont{Guinea}}, \bibnamefont{and}
  \bibinfo{author}{\bibfnamefont{A.}~\bibnamefont{Brataas}},
  \bibinfo{journal}{Phys. Rev. B.} \textbf{\bibinfo{volume}{74}},
  \bibinfo{pages}{155426} (\bibinfo{year}{2006}).

\bibitem[{\citenamefont{Odintsov and Yoshioka}(1999)}]{Odintsov99}
\bibinfo{author}{\bibfnamefont{A.}~\bibnamefont{Odintsov}} \bibnamefont{and}
  \bibinfo{author}{\bibfnamefont{H.}~\bibnamefont{Yoshioka}},
  \bibinfo{journal}{Phys. Rev. B.} \textbf{\bibinfo{volume}{59}},
  \bibinfo{pages}{10457} (\bibinfo{year}{1999}).

\bibitem[{\citenamefont{Ando}(2005)}]{Ando05}
\bibinfo{author}{\bibfnamefont{T.}~\bibnamefont{Ando}}, \bibinfo{journal}{J.
  Phys. Soc. Jpn.} \textbf{\bibinfo{volume}{74}}, \bibinfo{pages}{777}
  (\bibinfo{year}{2005}).

\bibitem[{\citenamefont{Minot et~al.}(2004)\citenamefont{Minot, Yaish,
  Sazonova, and McEuen}}]{Minot04}
\bibinfo{author}{\bibfnamefont{E.~D.} \bibnamefont{Minot}},
  \bibinfo{author}{\bibfnamefont{Y.}~\bibnamefont{Yaish}},
  \bibinfo{author}{\bibfnamefont{V.}~\bibnamefont{Sazonova}}, \bibnamefont{and}
  \bibinfo{author}{\bibfnamefont{P.~L.} \bibnamefont{McEuen}},
  \bibinfo{journal}{Nature} \textbf{\bibinfo{volume}{428}},
  \bibinfo{pages}{536} (\bibinfo{year}{2004}).

\bibitem[{\citenamefont{Abramowitz and Stegun}(1984)}]{abramowitz}
\bibinfo{author}{\bibfnamefont{M.}~\bibnamefont{Abramowitz}} \bibnamefont{and}
  \bibinfo{author}{\bibfnamefont{I.~A.} \bibnamefont{Stegun}},
  \emph{\bibinfo{title}{Pocketbook of mathematical functions}}
  (\bibinfo{publisher}{Harri Deutsch}, \bibinfo{year}{1984}).

\bibitem[{\citenamefont{Kostyrko and Krompiewski}(2008)}]{Kostyrko08}
\bibinfo{author}{\bibfnamefont{T.}~\bibnamefont{Kostyrko}} \bibnamefont{and}
  \bibinfo{author}{\bibfnamefont{S.}~\bibnamefont{Krompiewski}},
  \bibinfo{journal}{Semicond. Sci. Technol.} \textbf{\bibinfo{volume}{23}},
  \bibinfo{pages}{085024} (\bibinfo{year}{2008}).

\bibitem[{\citenamefont{Lieb and Mattis}(1962)}]{Lieb61}
\bibinfo{author}{\bibfnamefont{E.}~\bibnamefont{Lieb}} \bibnamefont{and}
  \bibinfo{author}{\bibfnamefont{D.}~\bibnamefont{Mattis}},
  \bibinfo{journal}{Phys. Rev.} \textbf{\bibinfo{volume}{125}},
  \bibinfo{pages}{164} (\bibinfo{year}{1962}).

\bibitem[{\citenamefont{Galland and Imamoglu}(2008)}]{Galland08}
\bibinfo{author}{\bibfnamefont{C.}~\bibnamefont{Galland}} \bibnamefont{and}
  \bibinfo{author}{\bibfnamefont{A.}~\bibnamefont{Imamoglu}},
  \bibinfo{journal}{Phys. Rev. Lett.} \textbf{\bibinfo{volume}{101}},
  \bibinfo{pages}{157404} (\bibinfo{year}{2008}).

\bibitem[{\citenamefont{Churchill et~al.}(2008)\citenamefont{Churchill,
  Kuemmeth, Harlow, Bestwick, Rashba, Flensberg, Stwertka, Taychatanapat,
  Watson, and Marcus}}]{Churchill08}
\bibinfo{author}{\bibfnamefont{H.}~\bibnamefont{Churchill}},
  \bibinfo{author}{\bibfnamefont{F.}~\bibnamefont{Kuemmeth}},
  \bibinfo{author}{\bibfnamefont{J.~W.} \bibnamefont{Harlow}},
  \bibinfo{author}{\bibfnamefont{A.~J.} \bibnamefont{Bestwick}},
  \bibinfo{author}{\bibfnamefont{E.~I.} \bibnamefont{Rashba}},
  \bibinfo{author}{\bibfnamefont{K.}~\bibnamefont{Flensberg}},
  \bibinfo{author}{\bibfnamefont{C.~H.} \bibnamefont{Stwertka}},
  \bibinfo{author}{\bibfnamefont{T.}~\bibnamefont{Taychatanapat}},
  \bibinfo{author}{\bibfnamefont{S.~K.} \bibnamefont{Watson}},
  \bibnamefont{and} \bibinfo{author}{\bibfnamefont{C.~M.}
  \bibnamefont{Marcus}}, \bibinfo{journal}{arXiv:0811.3239}
  (\bibinfo{year}{2008}).

\bibitem[{\citenamefont{Secchi and Rontani}(2009)}]{Secchi09}
\bibinfo{author}{\bibfnamefont{A.}~\bibnamefont{Secchi}} \bibnamefont{and}
  \bibinfo{author}{\bibfnamefont{M.}~\bibnamefont{Rontani}},
  \bibinfo{journal}{arXiv:0903.5107}  (\bibinfo{year}{2009}).

\end{thebibliography}

\end{document}